\documentstyle[12pt,aaspp4]{article}
\makeatletter
\def\lnyoro{\mathrel{\mathpalette\gl@align<}}
\def\gnyoro{\mathrel{\mathpalette\gl@align>}}
\def\gl@align#1#2{\lower.6ex\vbox{\baselineskip\z@skip\lineskip\z@\ialign{$\m@th
#1\hfil##\hfil$\crcr#2\crcr\sim\crcr}}}
\makeatother
\begin{document}

\title{\bf CYCLIC CHANGES OF DUST-TO-GAS RATIO}
\author{\bf HIROYUKI HIRASHITA$^{1,2}$}
\affil
{$^1$:  Department of Astronomy, Faculty of Science, Kyoto University,
Sakyo-ku, Kyoto 606-8502, Japan}
\affil
{$^2$:  Research Fellow of the Japan Society for the Promotion of
Science}
\centerline{accepted to {\sl ApJ}}
\centerline{email: hirasita@kusastro.kyoto-u.ac.jp}
\authoremail{hirasita@kusastro.kyoto-u.ac.jp}
\begin{abstract}

We discuss the time variation of dust-to-gas mass ratio  in 
spiral galaxies using the multi-phase model of interstellar
medium. The typical timescale of the phase change of an
interstellar gas is $\sim 10^{7\mbox{--}8}$ yr in spiral galaxies.
Since the phase transition changes the filling factor of
the cold gas where the dust growth occurs, the dust growth rate
varies on that timescale. In order to examine the response of
the dust-to-gas ratio to the phase transition, we construct a
model of the time evolution of the dust-to-gas ratio. We adopt
the three phase model for the interstellar gas and
the Ikeuchi-Tomita model for the mass exchange among the phases.
According to the model, three types of solutions are possible:
[1] all the gas is transformed to a hot gas;
[2] a stable stationary state of three phases is realized;
[3] the filling factor of each phase cyclically changes.
For each of the three types of solutions, the dust-to-gas
ratio behaves as follows:
[1] almost all the dust is destroyed (the dust-to-gas ratio
becomes $\sim 0$);
[2] the dust-to-gas ratio converges to a stationary state;
[3] the dust-to-gas ratio varies cyclically in response to the
phase transition.
In the case of [3], the amplitude of the variation of the
dust-to-gas ratio is large (nearly an order of magnitude) if
the dust growth timescale is shorter than the phase transition
timescale. This condition is easily satisfied in spiral galaxies.
However, it is difficult for dwarf galaxies to realize the
condition because their small metallicity makes the dust growth
timescale long.

\end{abstract}


\section{INTRODUCTION}

Some of the recent chemical evolution models of galaxies
have successfully included the dust
content (Wang 1991;
Lisenfeld \& Ferrara 1998; Dwek 1998;
Hirashita 1999a, hereafter
H99). Takagi, Arimoto, \& Vansevi\v{c}ius (1999) modeled the
dust extinction in ultraviolet (UV)--optical wavelengths and the
reprocessed light in the far-infrared (FIR) region, presenting
the spectral evolution in the UV--FIR wavelengths.

As to the dust formation,
supernovae (SNe) are known to be one of the most dominant
sources (Barlow 1978; Draine \& Salpeter 1979;
Dwek \& Scalo 1980), and SN
shocks destroy grains (McKee et al.\ 1987; Jones et al.\ 1994; 
Borkowski \& Dwek 1995). Thus,
dust content is connected with star formation histories.
In our previous work, H99, the dust-to-gas mass ratio was
expressed as a non-linear function of metallicity, which is
also related to star formation histories (see also
Lisenfeld \& Ferrara 1998). H99 confirmed the suggestion proposed
by  Dwek (1998) that the accretion process onto preexisting dust
grains is efficient in spiral galaxies. This is also supported by
the observations of elemental depletions in dense clouds
(e.g, Draine 1990). Since the accretion is effective in cold
clouds, the global efficiency of the accretion depends on the
fractional mass of the cold gas (Seab 1987; McKee 1989;
Draine 1990; Hirashita 1999b). Thus, the efficiency varies
on a timescale of the phase transition
of the interstellar medium (ISM;  $\sim 10^{7\mbox{--}8}$ yr;
Ikeuchi 1988; McKee 1989).

McKee \& Ostriker (1977) constructed a standard model for the ISM
composed of three phases: the hot rarefied gas
($T\sim 10^6$ K and $n\sim 10^{-3}$ cm$^{-3}$),
the warm gas ($T\sim 10^4$ K and $n\sim 10^{-1}$ cm$^{-3}$),
and the cold cloud ($T\sim 10^2$ K and $n\sim 10$ cm$^{-3}$).
One of the simplest models for the mass exchange
 among the
three ISM phases is described by Ikeuchi \& Tomita (1983,
hereafter IT83; see also Habe, Ikeuchi \& Tanaka 1981).
They included the following three processes: [1] the sweeping of
a warm gas into a cold component, [2] the evaporation of cold
clouds embedded in a hot gas, and [3] the radiative cooling
of a hot gas by collisions with a warm gas. Mass fraction
of each component can show an oscillatory behavior, which is
known as a limit-cycle (Nicolis \& Prigogine 1977;
see also Fujimoto \& Ikeuchi 1984 for another modeling of
interstellar gas). The limit-cycle behavior is supported by
Kamaya \& Takeuchi (1997, hereafter KT97),
who interpreted the various level of the star formation 
activities in spiral galaxies shown by
Tomita, Tomita, \& Sait\={o} (1996) using the limit-cycle model.

In the
limit-cycle model by IT83, mass fraction of each phase oscillates
continuously because of the mass exchange among the three
components of the ISM (see also
Korchagin, Ryabtsev, \& Vorobyov 1994).  The timescale of the
phase transition in the model is determined by three timescales
of the above physical processes [1]--[3]. Actually, a static
solution as well as the limit-cycle solution for the mass fractions
is possible for some range in the parameter space (this case is
also examined in this paper). However, the oscillatory behavior
(i.e., the limit-cycle model) of the fractional masses is supported
observationally. Indeed, the observed scatter of the FIR-to-optical
flux ratios of spiral galaxies (Tomita, Tomita, \& Sait\={o} 1996)
is interpreted through the limit-cycle model in KT97, who
suggested that the fractional mass of the cold phase changes in
the range of 0.1 to 0.7 (or more) on the timescale of
$10^{7\mbox{--}8}$ yr and that this leads to the time
variation of star formation activities if the Schmidt law
(Schmidt 1959) is applicable.

Hirashita (1999b)
combined the framework of H99 with a theoretical work on
multiphase ISM and
suggested the time variation of the dust-to-gas ratio by the phase
transition. We aim at constructing a model of the time evolution
of dust-to-gas ratio including the effect of the ISM phase
transition. The model in this paper enables us to simulate the
time variation of the
dust-to-gas ratio in a galaxy. This paper is organized as follows.
First of all, in the next section, a model for the time
variation of the dust-to-gas ratio is presented. Next, we
show the numerical solution of the model in \S 3. In \S 4, we
discuss the solution and present observational implications.
Finally, \S 5 is devoted to summary.

\section{MODELING OF DUST-TO-GAS RATIO AND ISM PHASE TRANSITION}

\subsection{Dust-to-Gas Ratio}

First, we prepare a model for the time variation of the
dust-to-gas ratio in a galaxy. The model is developed by H99
(see also Lisenfeld \& Ferrara 1998 and Dwek 1998) in order
to account for the relation between metallicity and dust-to-gas
ratio of nearby spiral galaxies. He adopted a simple
one-zone model (i.e., the galaxy is treated as a closed system and
spatial variations of physical quantities within the galaxy is
neglected). The model equations describe the changing rate of th
mass of gas, metal, and dust:
\begin{eqnarray}
\frac{dM_{\rm g}}{dt} & = & -\psi +E,\label{basic1} \\
\frac{dM_i}{dt}       & = & -X_i\psi +E_i,\label{basic2} \\
\frac{dM_{{\rm d},i}}{dt} & = & f_{{\rm in},i}E_i-\alpha f_iX_i\psi
+\frac{M_{{\rm d},i}(1-f_i)}{\tau_{\rm acc}}-
\frac{M_{{\rm d},i}}{\tau_{\rm SN}}.
\label{basic3}
\end{eqnarray}
Here, $M_{\rm g}$ is the mass of gas; $M_{i}$ and
$M_{{\rm d},i}$ denote the total mass of the metal $i$ ($i={\rm O}$,
C, Si, Mg, Fe, etc.) in both
gas and dust phases, and only in the dust
phase, respectively. The star formation rate is denoted by $\psi$;
$E$ is the total injection rate of
mass from stars; $X_i$ is the mass fraction of the element $i$
(i.e., $X_i\equiv M_i/M_{\rm g}$); $E_i$ is the total
injection rate of element $i$ from stars; $f_i$ is the mass fraction
of the element $i$ locked up in dust (i.e., $f_i=M_{{\rm d},i}/M_i$).
The meanings of the other parameters in the above equations are as
follows: $f_{{\rm in},i}$ represents the dust mass fraction in the
injected material (in other words, the dust condensation efficiency
in the ejecta such as stellar winds or SNe); $\alpha$ refers
to the efficiency of dust destruction during a star formation
[$\alpha =1$ corresponds to destruction of
only the dust incorporated into the star, and $\alpha >1\, (\alpha
<1)$ corresponds to a net destruction (formation) in the star
formation]; $\tau_{\rm acc}$ is the accretion timescale of the
element $i$ onto preexisting dust grains in molecular clouds;
$\tau_{\rm SN}$ is the timescale of dust destruction
by SN shocks. We hereafter assume that only dust
incorporated into stars is destroyed when the stars form; i.e.,
$\alpha =1$. The mass loss from the galaxy is neglected in this
paper, since we mainly treat spiral galaxies
and expect that outflows are prevented by their deep dark matter
potentials (Habe \& Ikeuchi 1980). When we consider star burst
phase (Habe \& Ikeuchi 1980) or dwarf galaxies
(Mac Low \& Ferrara 1999), the outflows may be important
for the dust content as well as gas and metal content
(Lisenfeld \& Ferrara 1998).

We adopt the instantaneous recycling approximation (i.e., a star
with the mass larger
than $m_{\rm l}$ dies instantaneously after its birth, leaving
the remnant of mass $w_m$). Once the initial mass function is
fixed, we can calculate the returned fraction of the mass
from formed stars (${\cal R}$)
and the mass fraction of the element $i$ that is newly
produced and ejected by stars (${\cal Y}_i$).
Using ${\cal R}$ and ${\cal Y}_i$, $E$ and $E_i$ in
equations (\ref{basic1})--(\ref{basic3}) is written as
(see H99 for the details)
\begin{eqnarray}
E & = & {\cal R}\psi , \\
E_i & = & ({\cal R}X_i+{\cal Y}_i)\psi .
\end{eqnarray}
After some algebra,
equations (\ref{basic2}) and (\ref{basic3}) becomes
\begin{eqnarray}
\tau_{\rm SF}\frac{dX_i}{dt} & = & {\cal Y}_i,
\label{basic5} \\
\tau_{\rm SF}\frac{d{\cal D}_i}{dt} & = & f_{{\rm in},i}
({\cal R}X_i+{\cal Y}_i)
-[\alpha -1+{\cal R}-\beta_{\rm acc}(1-f_i)+\beta_{\rm SN}]
{\cal D}_i,\label{basic6}
\end{eqnarray}
where $\tau_{\rm SF}\equiv M_{\rm g}/\psi$ (timescale of gas
consumption; e.g., Roberts 1963),
${\cal D}_i\equiv M_{{\rm d},i}/M_{\rm g}=f_iX_i$,
$\beta_{\rm acc}\equiv{\tau_{\rm SF}}/{\tau_{\rm acc}}$,
and $\beta_{\rm SN}\equiv{\tau_{\rm SF}}/{\tau_{\rm SN}}$.
The timescales above are estimated in the case of typical spiral
galaxies as $\tau_{\rm SF}\sim 1$--10 Gyr
(Kennicutt, Tamblyn, \& Congdon 1994), and
$\tau_{\rm acc}\sim\tau_{\rm SN}\sim 10^8$ yr (Draine 1990).

The two timescales, $\tau_{\rm acc}$ and $\tau_{\rm SN}$ depend
on the phase of the ISM where the dust exists
(Draine 1990; Hirashita 1999b). The accretion timescale,
$\tau_{\rm acc}$, is shorter in the cold phase of the ISM than in
any other phase, since efficient accretion needs a dense
environment. If we express the growth timescale of a dust grain in
the cold component as $\tau_{\rm grow}$, the accretion timescale is
expressed as $\tau_{\rm acc}={\tau_{\rm grow}}/{X_{\rm cold}}$,
where $X_{\rm cold}$ represents the mass fraction of the
cold phase to the total mass of ISM (Hirashita 1999b). This
is equivalent to
\begin{eqnarray}
\beta_{\rm acc}=X_{\rm cold}\beta_{\rm grow},\label{tacc}
\end{eqnarray}
where $\beta_{\rm grow}\equiv\tau_{\rm SF}/\tau_{\rm grow}$.
According to Draine (1990),
$\tau_{\rm grow}\sim 5\times 10^7$ yr in the cold gas.
The destruction timescale of the dust $\beta_{\rm SN}$ is
kept constant for simplicity, since
we would like to concentrate on the time variation of the dust growth
efficiency to examine the result in Hirashita (1999b).
The star formation history should be taken into account in order to
model the time dependence of $\beta_{\rm SN}$. Moreover, the time
variation of the SN rate
also contributes to the time variation of the fractional masses the
three ISM components. Because of such complexities
in the time dependence of $\beta_{\rm SN}$, we simply treat
$\beta_{\rm SN}$ as constant.

\subsection{ISM Phase Transition}

Next, we review the model by IT83. This model is used to
calculate the time evolution of the
filling factors of the three ISM phases (see also \S 1).
The result is used to calculate the dust formation efficiency
through equation (\ref{tacc}).

The interstellar medium is assumed to consist of three
components (McKee \& Ostriker 1977); the hot rarefied gas
($T\sim 10^6$ K, $n\sim 10^{-3}$ cm$^{-3}$), the warm gas
($T\sim 10^4$ K, $n\sim 10^{-1}$ cm$^{-3}$), and the cold
cloud ($T\sim 10^2$ K, $n\sim 10$ cm$^{-3}$).
The fractional masses of the three components are
$X_{\rm hot}$, $X_{\rm warm}$, and $X_{\rm cold}$, respectively.
A trivial relation is
\begin{eqnarray}
X_{\rm hot}+X_{\rm warm}+X_{\rm cold}=1.\label{trivial}
\end{eqnarray}
The following three processes are considered
(see IT83 for the details):
[1] the sweeping of a warm gas into a cold component at the
rate of $aX_{\rm warm}$ ($a\sim 5\times 10^{-8}$ yr$^{-1}$);
[2] the evaporation of cold clouds
embedded in a hot gas at the rate of $bX_{\rm cold}X_{\rm hot}^2$
($b\sim 10^{-7}$--$10^{-8}$ yr$^{-1}$);
[3] the radiative cooling of a hot gas through collisions with
a warm gas at the rate of $cX_{\rm warm}X_{\rm hot}$
($c\sim 10^{-6}$--$10^{-7}$ yr$^{-1}$). Writing down
the rate equations and using equation (\ref{trivial}),
we obtain
\begin{eqnarray}
\frac{dX_{\rm cold}}{d\tau} & = & -BX_{\rm cold}X_{\rm hot}^2+
A(1-X_{\rm cold}-X_{\rm hot}),\label{cold} \\
\frac{dX_{\rm hot}}{d\tau} & = & -X_{\rm hot}
(1-X_{\rm cold}-X_{\rm hot})+BX_{\rm cold}X_{\rm hot}^2,
\label{hot}
\end{eqnarray}
where $\tau\equiv ct$, $A\equiv a/c$, and $B\equiv b/c$.

The solutions of equations (\ref{cold}) and (\ref{hot}) are
classified into the following three types (IT83):

[1] $A>1$; all the orbits in the
$(X_{\rm cold},\; X_{\rm hot})$-plane
reduce to the node (0, 1),

[2] $A<1$ and $B>B_{\rm cr}$; all the orbits reduce to a stable
focus $[(1-A)/(AB+1),\; A]$,

[3] $A<1$ and $B<B_{\rm cr}$; all the orbits converge on a
limit-cycle orbit,

\noindent
where $B_{\rm cr}\equiv (1-2A)/A^2$.
As a representative parameter sets for [1] and [2], we examine
$(A,\; B)=(1.2,\; 3.0)$ and $(A,\; B)=(0.4,\; 3.0)$,
respectively (denoted as Models 1 and 2 in Table 1,
respectively). We especially concentrate
on the case [3], since we treat spiral galaxies with
cold gas and with various levels of star formation activities
(e.g., Tomita et al.\ 1996).
The two properties of spiral galaxies are naturally realized
 (KT97) if the filling factor of the cold gas shows the
limit-cycle behavior and the star formation rate
changes according to the Schmidt law
($\psi\propto X_{\rm cold}^n$; $n=1$--2; Schmidt 1959).
As representative parameter sets of the limit-cycle orbit,
we examine the case of $(A,\; B)=(0.3,\; 3.0)$ and
$(A,\; B)=(0.3,\; 0.5)$
(denoted by Model 3 and 4 in Table 1, respectively)
\footnote{Two cases for each model are examined for different
values of $c$ as explained in \S 2.3 (Model 3a, 3b, 4a, and 4b).}.
The cycle periods of the former case is about three times
shorter than that of the latter.

In Table 1, the adopted parameters are summarized.
The column ``type'' means the behavior of the solutions
and named ``runaway,'' ``stationary'' and ``limit-cycle''
for the cases [1], [2] and [3], respectively.
We note that all the parameter sets above are examined in 
 IT83 (their Figures 1 and 3).

\subsection{Summary of the Basic Equations}

Here, we summarize the equations and the parameters.

First, the filling factor of the cold gas is calculated by
the following set of equations:
\begin{eqnarray}
\frac{dX_{\rm cold}}{d\tau} & = & -BX_{\rm cold}X_{\rm hot}^2+
A(1-X_{\rm cold}-X_{\rm hot}),\label{basic8} \\
\frac{dX_{\rm hot}}{d\tau} & = & -X_{\rm hot}
(1-X_{\rm cold}-X_{\rm hot})+BX_{\rm cold}X_{\rm hot}^2,
\label{basic9}
\end{eqnarray}
with parameters $A$ and $B$ listed in Table 1. According to
these equations, the time variation of the cold-gas fractional
mass $X_{\rm cold}$ is calculated to obtain the efficiency
of the heavy-element
accretion onto dust grains (eq.~\ref{tacc}).

Next, the time variation of the dust-to-gas ratio is
simultaneously solved. Since we are interested in a timescale
much shorter than $\tau_{\rm SF}$, which is also interpreted
as metal-production timescale through equation (\ref{basic5}),
we treat $X_i$ as constants. Then, equation (\ref{basic6}) is
solved for oxygen abundance ($i={\rm O}$, because we adopted the
oxygen as a tracer element in the previous work, H99: see also
Lisenfeld \& Ferrara 1998):
\begin{eqnarray}
c\tau_{\rm SF}\frac{d{\cal D}_{\rm O}}{d\tau} =
f_{{\rm in,\, O}}({\cal R}X_{\rm O}+{\cal Y}_{\rm O})
-[{\cal R}-\beta_{\rm acc}(1-f_{\rm O})+\beta_{\rm SN}]
{\cal D}_{\rm O},\label{basic7}
\end{eqnarray}
where we express the time in a nondimensional manner by using
$\tau =ct$ and we assume $\alpha =1$ (equivalent to the assumption
that only dust incorporated into stars is destroyed in the star
formation processes; \S 2.1). We note the relation
$f_{\rm O}={\cal D}_{\rm O}/X_{\rm O}$. We fix the parameters as
$\tau_{\rm SF}= 3\times10^9$ yr (e.g., Kennicutt et al.\ 1994),
$f_{{\rm in,\, O}}=0.05$ (H99), ${\cal R}=0.79$,
${\cal Y}_{\rm O}=1.8\times 10^{-2}$,
$\beta_{\rm grow}=100$ (corresponding to
$\tau_{\rm grow}=\tau_{\rm SF}/100=3\times 10^7$ yr; we note
that $\beta_{\rm grow}$ and $\beta_{\rm acc}$ are related through
eq.\ \ref{tacc}),
$\beta_{\rm SN}=10$ (H99), and $X_{\rm O}=0.013$
(the solar system abundance).
The dependence on these parameters above are described in
Lisenfeld \& Ferrara (1998), H99, and Hirashita (1999c)
except for $c$ and $\tau_{\rm SF}$. The parameter $c$ scales
the phase transition timescale, and we examine the cases of
$c\tau_{\rm SF}=100$ and 1000 (Table 1:
$c^{-1}=3\times 10^6$--$3\times 10^7$ yr; IT83). 
We note that $c$ appears in the form of the nondimensional parameter
$c\tau_{\rm SF}$. We examine
only the case of $c\tau_{\rm SF}=100$ in Model 1 and 2, since
the qualitative behavior of the dust-to-gas ratio is the same,
irrespective of which value of $c\tau_{\rm SF}$ is adopted.

Finally, the total dust-to-gas ratio ${\cal D}$ is calculated
from the following equation:
\begin{eqnarray}
{\cal D}=C{\cal D}_{\rm O},\label{conversion}
\end{eqnarray}
where we assume $C=2.2$ (the Galactic value; Whittet 1992,
p.\ 52).

We solve the basic equations (\ref{basic8})--(\ref{conversion})
for the parameters listed in Table 1.
The initial condition for the filling factors are set according
to IT83:
$(X_{\rm cold},\; X_{\rm hot})=(0.0,\; 0.1)$ except for the Model
4 and
$(X_{\rm cold},\; X_{\rm hot})=(0.0,\; 0.7)$ for Model 4.
The initial dust-to-gas ratio is set as
${\cal D}=0.0080$ (Galactic value; Whittet 1992) for all the
models. Models 3 and 4 are examined for two different values
of $c$ listed in Table 1.

\section{RESULTS}

First, the result of Model 1 is displayed in Figure 1. Because the
cold gas disappears, the dust growth becomes ineffective. Then the
dust destruction makes the dust-to-gas ratio converge to $\sim 0$
in the dust destruction timescale
($\sim 3\times 10^8$ yr, i.e., 10 in the unit of $c^{-1}$).
This model represents the efficient sweeping of warm gas
in comparison with the radiative cooling of the hot gas.
Thus, the swept gas is first converted into the cold component
and then into hot gas (IT83). If the hot gas escapes out of the
galaxy owing to its large thermal energy, this model corresponds
to the galactic wind model of elliptical galaxies (IT83;
Arimoto \& Yoshii 1987).

The final value of the dust-to-gas ratio in Model 1 is
calculated by considering a stationary state,
$d{\cal D}_{\rm O}/dt=0$, in equation (\ref{basic7}):
\begin{eqnarray}
\frac{\beta_{\rm acc}}{X_{\rm O}}{\cal D}_{\rm O}^2+
({\cal R}-\beta_{\rm acc}+\beta_{\rm SN}){\cal D}_{\rm O}-
f_{\rm in,\, O}({\cal R}X_{\rm O}+{\cal Y}_{\rm O})=0.
\label{stationary}
\end{eqnarray}
Because of the constant dust formation from metal injection
at the rate of
$f_{\rm in,\, O}({\cal R}X_{\rm O}+{\cal Y}_{\rm O})$,
the dust-to-gas ratio is not exactly 0 even after a long time.
Since $X_{\rm cold}=0$, $\beta_{\rm acc}=0$.
Thus, the solution of equation(\ref{stationary}) is
\begin{eqnarray}
{\cal D}=C{\cal D}_{\rm O}=
\frac{Cf_{\rm in\, O}({\cal R}X_{\rm O}+{\cal Y}_{\rm O})}
{{\cal R}+\beta_{\rm SN}}=3\times 10^{-4},
\end{eqnarray}
 for the adopted parameters $X_{\rm O}=0.013$,
$f_{\rm in\, O}=0.05$, ${\cal R}=0.79$,
${\cal Y}_{\rm O}=0.018$, and $\beta_{\rm SN}=10$.
This indeed matches the final value of the dust-to-gas ratio
in Model 1.

Next, the result of Model 2 is presented in Figure 2. As shown
in IT83, the solution for $X_{\rm cold}$ reduces to a stable
point $(1-A)/(AB+1)=0.27$. The stable stationary state is
realized at a finite dust-to-gas ratio, which is determined so
that the dust formation rate is balanced with the dust destruction
rate. The value is calculated from equation (\ref{basic7}).
Putting the adopted parameters, the positive solution is
${\cal D}=C{\cal D}_{\rm O}\simeq 0.017$, which matches the
stationary value in Figure 2.

Finally, the limit-cycle case is examined in Figure 3 and 4.
The four sets of parameters listed in Table 1 are examined
(Models 3a, 3b, 4a, and 4b). The resulting time evolution of the
cold component of ISM
and the dust-to-gas ratio is presented in Figures 3a, 3b, 4a, and
4b for the four cases. As expected, the dust-to-gas ratio
oscillates in response to the fractional mass of the cold
component, though the amplitude of the variation largely depends
on the parameters. Comparing the Figures 3a and 4a
(or Figures 3b and 4b), we see that
the amplitude becomes larger if the period of the
oscillation is longer. This is because the dust has enough
time to grow if the period is long.
Also from the comparison between the Figures
3a and 3b (or Figures 4a and 4b), we find that the
smaller $c$ makes the amplitude larger. Since
small $c$ means that a timestep of the phase transition
model (eqs.\ \ref{cold} and \ref{hot}) is long in the real
time unit ($\tau =1$ corresponds to $t=c^{-1}$),
the dust has enough time to grow in the case of smaller $c$.
The timescale of the dust growth relative to that of the
phase transition determines the amplitude. This point
is again discussed in the next section.

\section{DISCUSSIONS}

\subsection{Timescales}

In the runaway and stationary solutions (Models 1 and 2),
the  dust-to-gas ratio is settled into a stationary value
determined by equation (\ref{stationary}) after the dust
formation and destruction timescales ($\sim 10^8$ yr).
In the limit-cycle type solutions (Models 3 and 4),
the amplitude of the dust-to-gas oscillation largely
depends on
the timescale of the phase transition. In this subsection,
we discuss the phase-transition and dust-formation
timescales in the limit-cycle model.

From the solution of the model equations in the previous
section (Figs.~3 and 4), we find that the dust-to-gas ratio
oscillates in response to the filling factor of the cold
component. However, the amplitude of the oscillation
is highly dependent on the values of the parameters
$A$, $B$, and $c$. In other words, the amplitude is determined
by the relation between the timescale of the phase transition
(or the period of the oscillation) and that of the
dust growth timescale. In the following, we discuss the relation
between the the period of the phase transition
(denoted by $\tau_{\rm tr}$) and the dust growth timescale
($\tau_{\rm grow}$).

In order for the dust to grow,
 the dust must have
enough time for the growth, i.e., $\tau_{\rm acc}<\tau_{\rm tr}$.
By using equation (\ref{tacc}), the condition is equivalent to
\begin{eqnarray}
\frac{\tau_{\rm grow}}{X_{\rm cold}}<\tau_{\rm tr},
\end{eqnarray}
which reads the following inequality by using 
 $\beta_{\rm grow}$:
\begin{eqnarray}
\frac{\tau_{\rm tr}}{\tau_{\rm SF}}>
\frac{1}{\beta_{\rm grow}X_{\rm cold}}.\label{cond1}
\end{eqnarray}
Putting $\beta_{\rm grow}=100$ and $X_{\rm cold}=0.5$,
we obtain ${\tau_{\rm tr}}/{\tau_{\rm SF}}>1/50$ for the enough
dust growth, which realizes the large amplitude of
the time variation of the dust-to-gas ratio. This means that
the timescale of the phase transition should be larger than
$\mbox{several}\times 10^7$ yr for the sufficient dust growth,
since $\tau_{\rm SF}\sim 3$ Gyr (Kennicutt et al.\ 1994).
Here, we note that the timestep in Figures 3 and 4 is $c^{-1}$ in
the physical unit and that
$c^{-1}=3\times 10^6$--$3\times 10^7$ yr. In the longer
unit ($c^{-1}=3\times 10^7$ yr; corresponding to
Figures 3a and 4a), the amplitude of the dust-to-gas-ratio
oscillation is large. On the contrary, in the shorter unit
($c^{-1}=3\times 10^6$ yr; corresponding to Figures 3b and 4b),
the amplitude is much smaller because the timestep is shorter
and thus the condition (\ref{cond1}) is difficult to be satisfied.
Moreover, comparing Figures 3a and 4a, we see that the amplitude is
larger if the period is longer.

In Figure 4a, the maximum of the
dust-to-gas ratio is an order of magnitude larger than the minimum.
Thus, the dust-to-gas ratio in spiral galaxies can show an
order-of-magnitude oscillation. We should note that the parameter
range is reasonable for the spiral galaxies (Ikeuchi 1988). Thus,
we have confirmed the interpretation of the scatter of the
dust-to-gas ratio in Hirashita (1999b), who interpreted the scatter
of the dust-to-gas ratio of the nearby spiral sample as the
short-term ($\sim 10^7$--$10^8$-yr) variation of the dust-to-gas
ratio with the maximum-to-minimum ratio of more than 4.

The important point is that the dust-to-gas-ratio oscillation
with large amplitude occurs in a reasonable parameter range for
spiral galaxies (Model 4a). However, the amplitudes are small
for the other models after the stable limit cycle is attained.
Thus, some spiral galaxies may not experience the
prominent oscillation of the dust-to-gas ratio.
But it is possible for the spiral galaxies to experience the
oscillation of the dust-to-gas ratio, which is observed
in the scatter of the dust-to-gas ratio of the spiral sample.

\subsection{Effect of Chemical Evolution}

In the discussions above,
the metallicity was fixed because the
timescale of the phase transition is much shorter than that of
the chemical enrichment. However, the time range shown in
the figures in this paper (250 timesteps) corresponds to 7.5 Gyr
when the $c\tau_{\rm SF}=100$ is adopted. Thus, we need to
examine the effect of the chemical evolution here.

To include the effect of chemical evolution, we solve
equation (\ref{basic5}) to examine the time variation of metallicity.
If $\tau_{\rm SF}$ is fixed for simplicity, the solution of equation
(\ref{basic5}) becomes
\begin{eqnarray}
X_{\rm O}=\frac{{\cal Y}_{\rm O}\tau}{c\tau_{\rm SF}}.\label{varx}
\end{eqnarray}
In other words, we simply assume a constant rate of metal
enrichment to examine a effect of the chemical evolution
qualitatively. We solve basic equations in \S 2.3 for
the time-variable $X_{\rm O}$ described by equation (\ref{varx}).
As a representative case, we adopt the parameter set identical to
Model 4a in Table1, since the oscillation behavior of
the dust-to-gas ratio is the most prominent (Figure 4a).
The result is shown in Figure 5. We see that the qualitative
behavior of the dust-to-gas ratio is the same as the
results for constant metallicity, except for the
gradual increase of dust-to-gas ratio. This increase is
due to the metal production, from which dust is formed.

Thus, two behaviors of the temporal variation is coupled:
One is the gradual increase of the dust-to-gas ratio on the
timescale of chemical enrichment, and the other is the 
short-term variation of the dust-to-gas ratio owing to
the ISM phase changes.

We note that the effect of chemical evolution is of
significant importance if $\tau_{\rm SF}(\sim$ timescale
of the chemical enrichment) is short. Especially for
starburst galaxies, $\tau_{\rm SF}\sim 10^8$ yr, which is
comparable to $\tau_{\rm tr}$ (Moorwood 1996). In this case,
the metal enrichment also proceeds from the starburst
in $\tau_{\rm SF}$, and thus the assumption of the constant
metallicity is never satisfied.
The same discussion may also be applied to the initial burst
in the formation epoch of galaxies, when $\tau_{\rm SF}$ is
nearly the value of starburst galaxies ($\sim 10^8$ yr;
the dynamical time).

\subsection{Observational Implications}

We have shown that the dust-to-gas ratio varies on a timescale
of the ISM phase transition ($\sim 10^7$--$10^8$ yr)
in spiral galaxies. The amplitude of the variation can be an
order of magnitude (Figs.\ 3a--d). This confirms the
previous suggestion in Hirashita (1999b) that the
dust-to-gas ratio scatter of the nearby spiral galaxies
are explained by the time variation. The result also suggests
that the variation is partially responsible for the
scatter of the FIR-to-optical flux ratio of spiral sample in
Tomita, Tomita, \& Sait\={o} (1996), since the
dust content as well as the stellar heating (dust temperature) is
responsible for the far-infrared luminosity
(e.g., Whittet 1992, p.\ 170).

We comment on the dust content in dwarf
galaxies. Lisenfeld \& Ferrara (1998) modeled the time
evolution of the dust amount in star-forming dwarf galaxies
(dwarf irregular galaxies and blue compact dwarf galaxies).
Based on their model and considering the dust formation in
cold clouds (Dwek 1998), Hirashita (1999c) presented that the
dust growth in dwarf galaxies is negligible compared with
the dust condensation from metal ejected by stars.
This is because the metallicity of dwarf galaxies are
much smaller than that of spiral galaxies and the collision
between grains and metal atoms are not frequent enough
for efficienct growth of the grains.
Thus, the oscillation of dust-to-gas
ratio through the variation of the dust growth efficiency
is difficult in dwarf galaxies. In this case the scatter of the
dust-to-gas ratio of the dwarf sample may be interpreted to
reflect the various efficiency of the gas outflow
(Lisenfeld \& Ferrara 1998). The wind may easily blows out
of the dwarf galaxies because of their shallow gravitational
potentials (Larson 1974; De Young \& Heckman 1994), though the
distribution of dark matter largely affects the process of the
outflow (Mac Low \& Ferrara 1999;
Ferrara \& Tolstoy 1999).

We should note that the timescale of the phase
transition in dwarf galaxies is expected to be longer because
of their longer cooling timescale. Since the metal-line
cooling is the dominant cooling mechanism for the hot gas
(Raymond, Cox, \& Smith 1976), the small metallicity of dwarf
galaxies, typically $\sim 1/10$ of the spiral galaxies,  
makes the cooling time an order of magnitude longer
($> 10^8$ yr). Thus, once the hot component becomes
dominant in a dwarf galaxy, the dwarf galaxy must wait 
for the hot component to cool down for more than
$10^8$ yr, during which the growth of dust grains is
prevented. This as well as the mechanism in the previous
paragraph makes the dust growth difficult.
Indeed, the dust-to-gas ratios in dwarf galaxies 
are known to be much smaller than those in spiral galaxies
(e.g., Gondhalekar et al.\ 1986).

\section{SUMMARY}

We have discussed the time evolution of dust-to-gas mass ratio
in the context of multi-phase model of interstellar medium in
spiral galaxies. The
phase transition of interstellar gas occurs on a timescale of
$\sim 10^7$--$10^8$ yr in spiral galaxies (IT83). The dust growth
rate also varies on that timescale owing to the time variation
of the fractional mass of the
cold gas through the phase transition (Hirashita 1999b;
\S 2.1 of this paper).
In order to examine the response of the dust-to-gas ratio to the
phase transition, we modeled the time variation of the
dust-to-gas ratio with a model of the phase transition.
 We adopt
the three phase model for the interstellar gas and
the Ikeuchi-Tomita model for the phase transition. According
to the model, three types of solutions are possible:
[1] all the gas is transformed to a hot gas;
[2] a stable stationary state of three phases is realized;
[3] the filling factors of each phase cyclically changes.
For each of the three types of solutions, the dust-to-gas
ratio behaves as follows (\S 3):
[1] the dust is fully destroyed (the dust-to-gas ratio becomes 0);
[2] the dust-to-gas ratio converges to a stationary state;
[3] the dust-to-gas ratio varies cyclically in response to the
phase transition.
When the case [3] is applicable, the amplitude of the
variation of the dust-to-gas ratio is large (nearly an order of
magnitude) if the dust growth timescale is shorter than
the phase transition timescale. This condition is satisfied
within the reasonable parameter range of spiral galaxies
(\S 4.1).

However, the model needs to be modified for dwarf galaxies,
since their small metallicity makes the dust growth rate in clouds
small (Hirashita 1999c),
and their timescale of the phase transition
is longer owing to their longer cooling timescale.
For starburst galaxies, the effect of chemical enrichment is
important because of their short metal-enrichment timescale.

\acknowledgements

We wish to thank the anonymous referee for useful comments
that substantially improved this paper.
We are grateful to S. Mineshige for his continuous encouragement. 
This work is supported by the Research Fellowship of the Japan
Society for the Promotion of Science for Young Scientists.
We have made extensive use
of the NASA's Astrophysics Data System Abstract Service (ADS).

\newpage

\begin{table}
\caption{Examined Models}
\begin{tabular}{ccccc}
\tableline\tableline
Model No. & $A$ & $B$ & $c\tau_{\rm SF}$ & type \\
\tableline
1    & 1.2 & 3.0 & 100  & runaway     \\
2    & 0.4 & 3.0 & 100  & stationary  \\
3a & 0.3 & 3.0 & 100  & limit-cycle \\
3b & 0.3 & 3.0 & 1000 & limit-cycle \\ 
4a & 0.3 & 0.5 & 100  & limit-cycle \\
4b & 0.3 & 0.5 & 1000 & limit-cycle \\\tableline
\end{tabular}
\end{table}

\centerline{\bf FIGURE CAPTIONS}

\noindent
FIG. 1---
{\it Upper panel}: the time evolution of the
the cold-gas mass filling factor (${X}_{\rm cold}$)
for Model 1. The initial condition is
$(X_{\rm cold},\; X_{\rm hot})=(0.0,\; 0.1)$.
The adopted values of the parameters are
$A=1.2$, $B =3.0$, and $c\tau_{\rm SF} =100$.
The timestep is normalized by $c^{-1}\simeq 3\times 10^7$ yr.
{\it Lower panel}: the time evolution of the dust-to-gas ratio
(${\cal D}$).

\medskip

\noindent
FIG. 2 ---
The same as Fig.\ 1, but for Model 2.
The initial condition is
$(X_{\rm cold},\; X_{\rm hot})=(0.0,\; 0.1)$.
The adopted values of the parameters are
$A=0.4$, $B =3.0$, and $c\tau_{\rm SF} =100$.
The timestep is normalized by $c^{-1}\simeq 3\times 10^7$ yr.

\medskip

\noindent
FIG. 3a ---
The same as Fig.\ 1, but for Model 3a.
The initial condition is
$(X_{\rm cold},\; X_{\rm hot})=(0.0,\; 0.1)$.
The adopted values of the parameters are
$A=0.3$, $B =3.0$, and $c\tau_{\rm SF} =100$.
The timestep is normalized by $c^{-1}\simeq 3\times 10^7$ yr.

\medskip

\noindent
FIG. 3b ---
The same as Fig.\ 1, but for Model 3b.
The initial condition is
$(X_{\rm cold},\; X_{\rm hot})=(0.0,\; 0.1)$.
The adopted values of the parameters are
$A=0.3$, $B =3.0$, and $c\tau_{\rm SF} =1000$.
The timestep is normalized by $c^{-1}\simeq 3\times 10^6$ yr.

\noindent
FIG. 4a ---
The same as Fig.\ 1, but for Model 4a.
The initial condition is
$(X_{\rm cold},\; X_{\rm hot})=(0.0,\; 0.7)$.
The adopted values of the parameters are
$A=0.3$, $B =0.5$, and $c\tau_{\rm SF} =100$.
The timestep is normalized by $c^{-1}\simeq 3\times 10^7$ yr.

\noindent
FIG. 4b ---
The same as Fig.\ 1, but for Model 4b.
The initial condition is
$(X_{\rm cold},\; X_{\rm hot})=(0.0,\; 0.7)$.
The adopted values of the parameters are
$A=0.3$, $B =0.5$, and $c\tau_{\rm SF} =1000$.
The timestep is normalized by $c^{-1}\simeq 3\times 10^6$ yr.

\noindent
FIG. 5 ---
Time evolution of the dust-to-gas ratio. The model parameters
are same as Fig.~4a, but the effect of the chemical evolution
is included. A constant rate of metal enrichment is assumed.

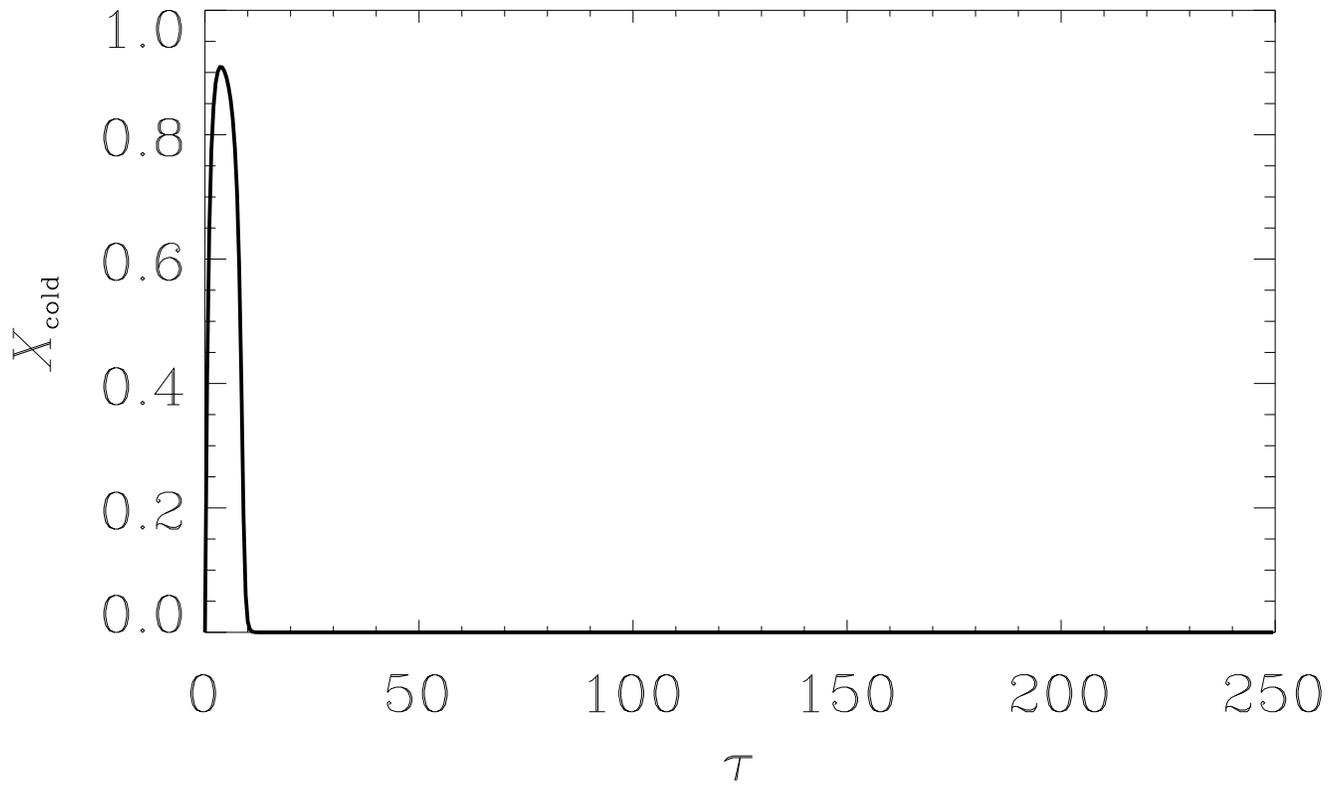
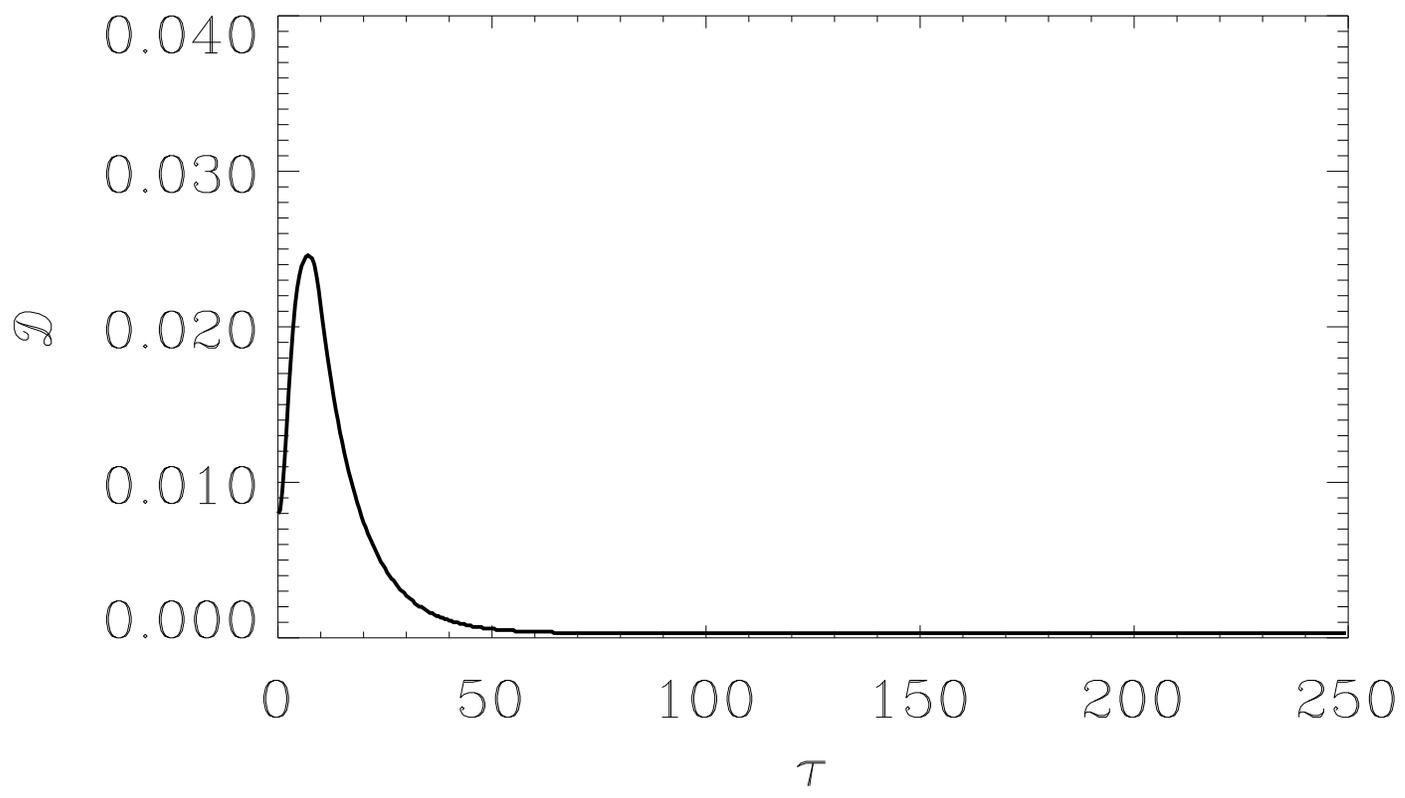
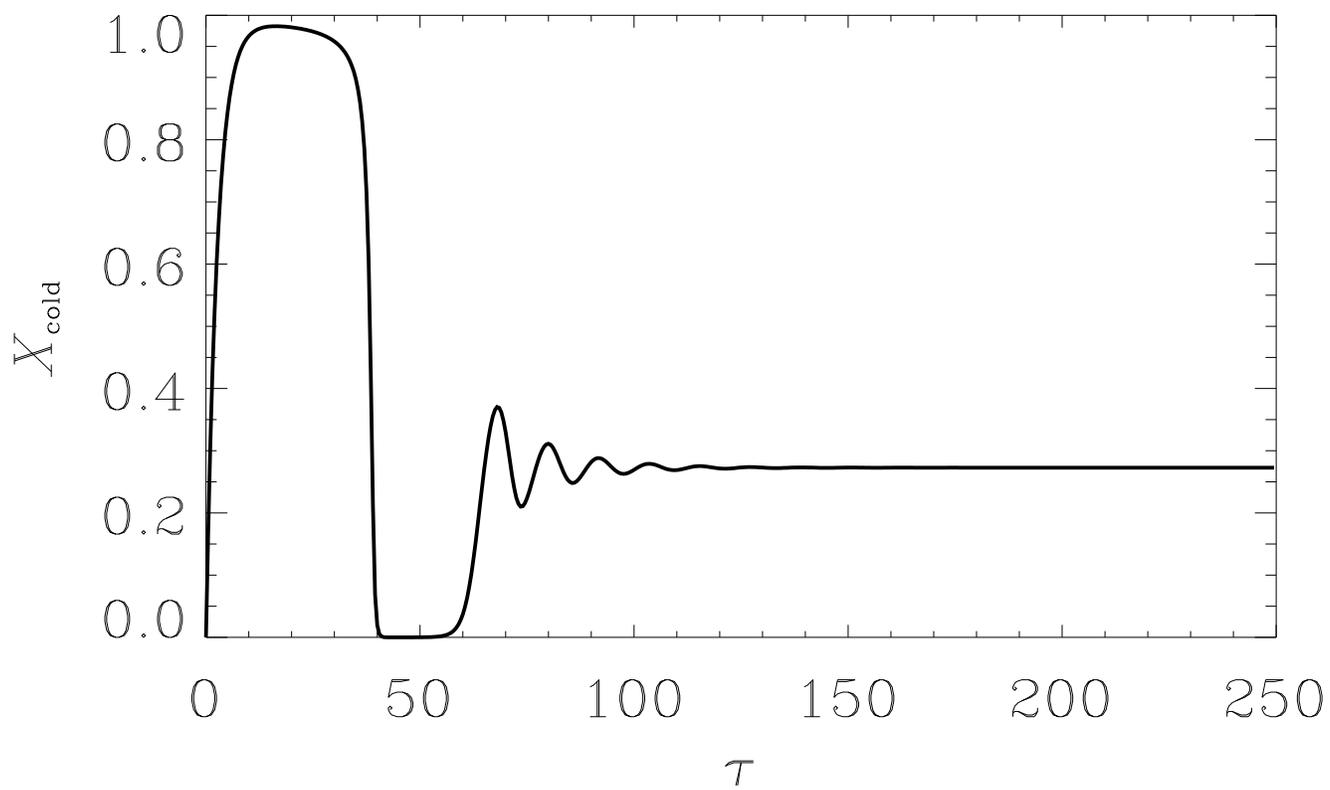
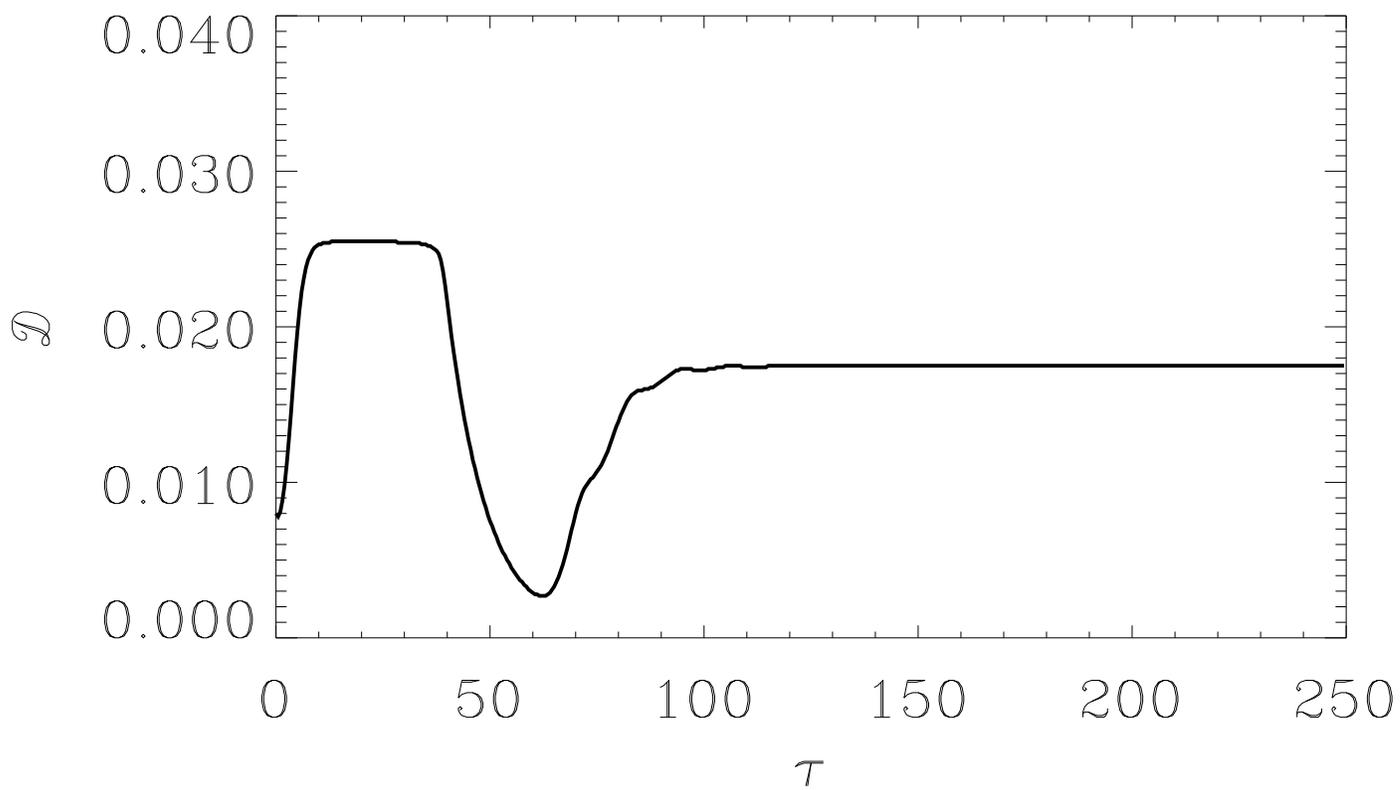
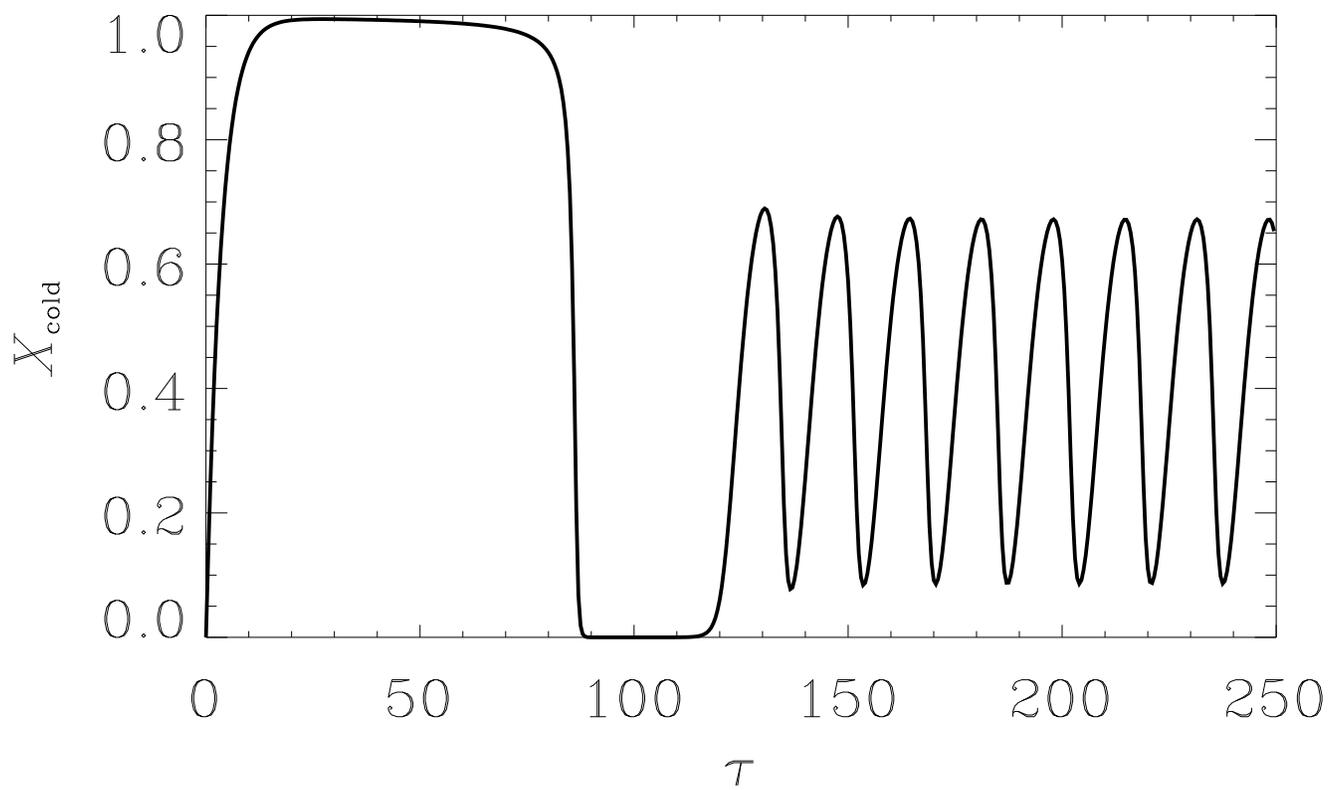
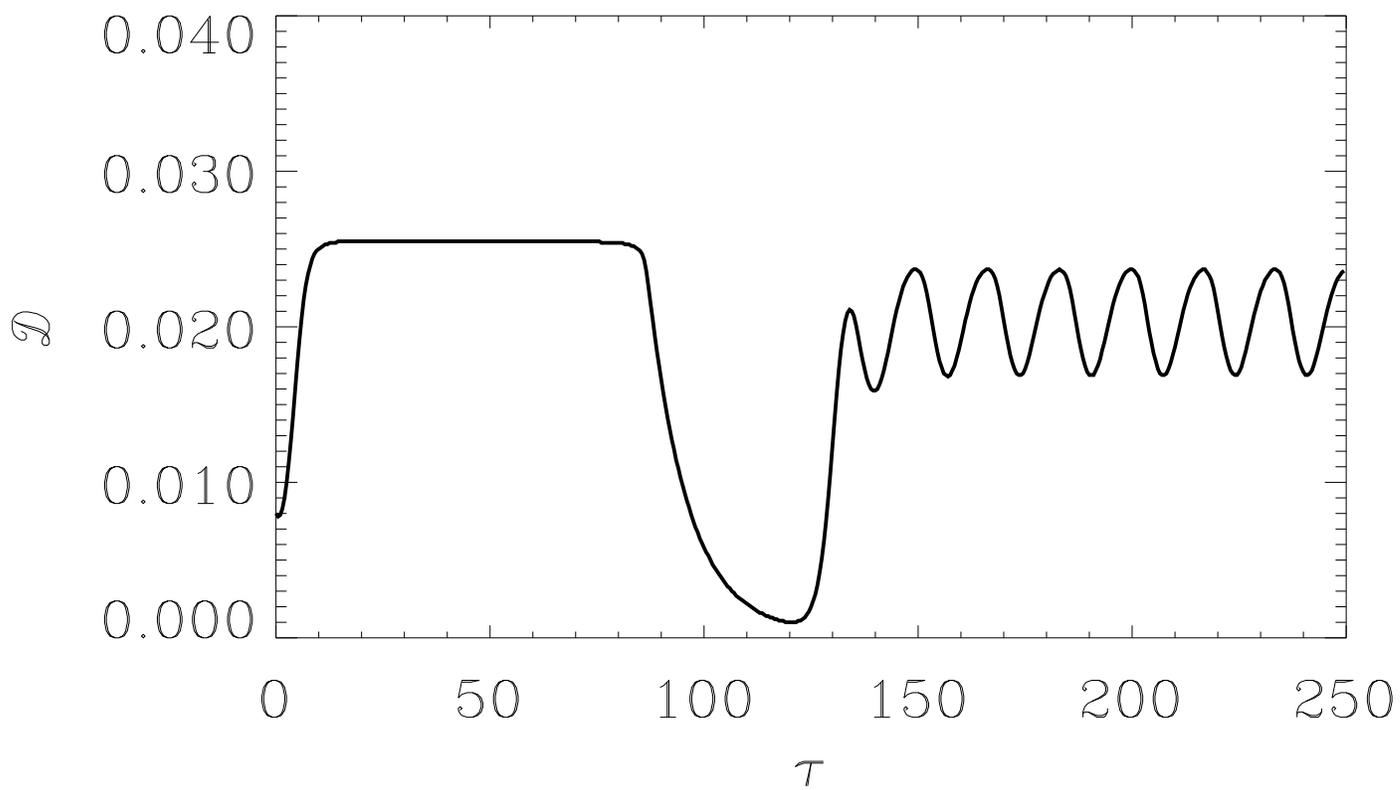
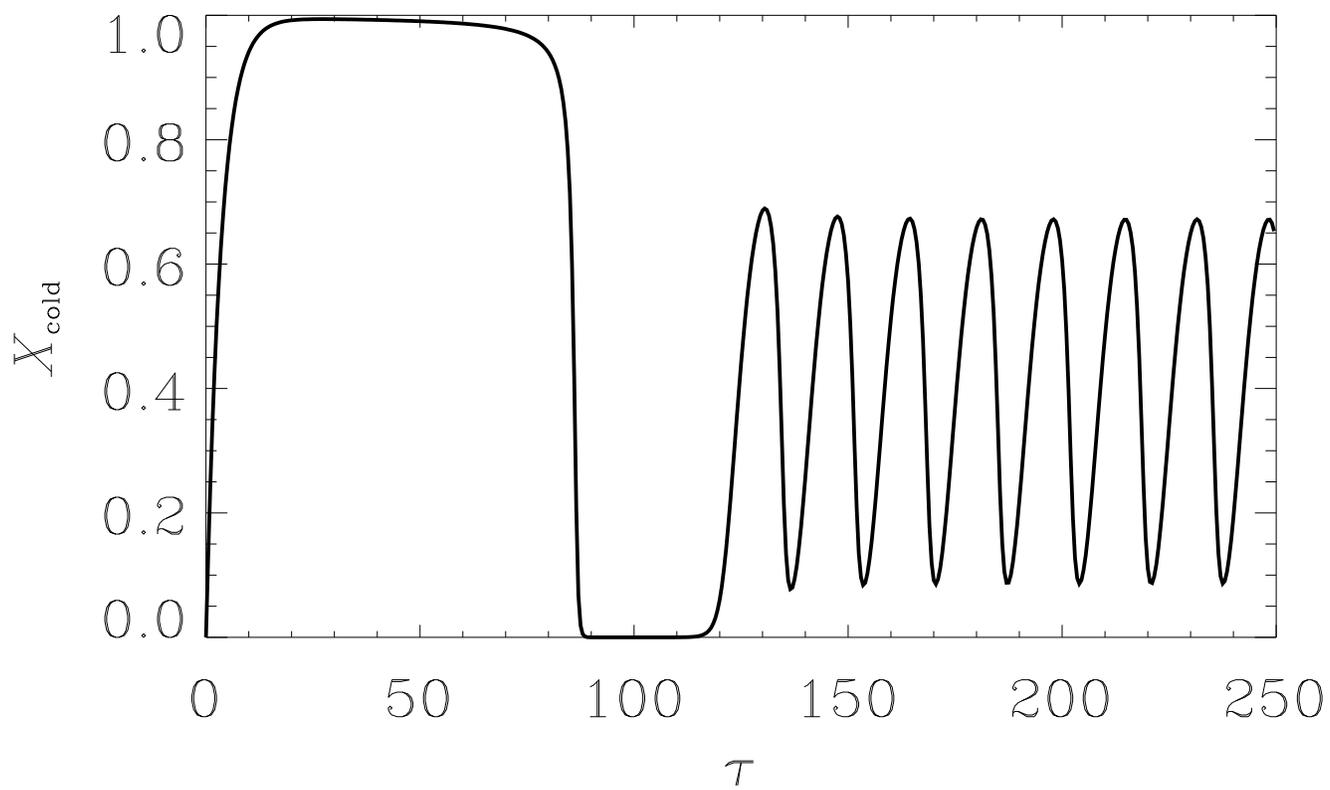
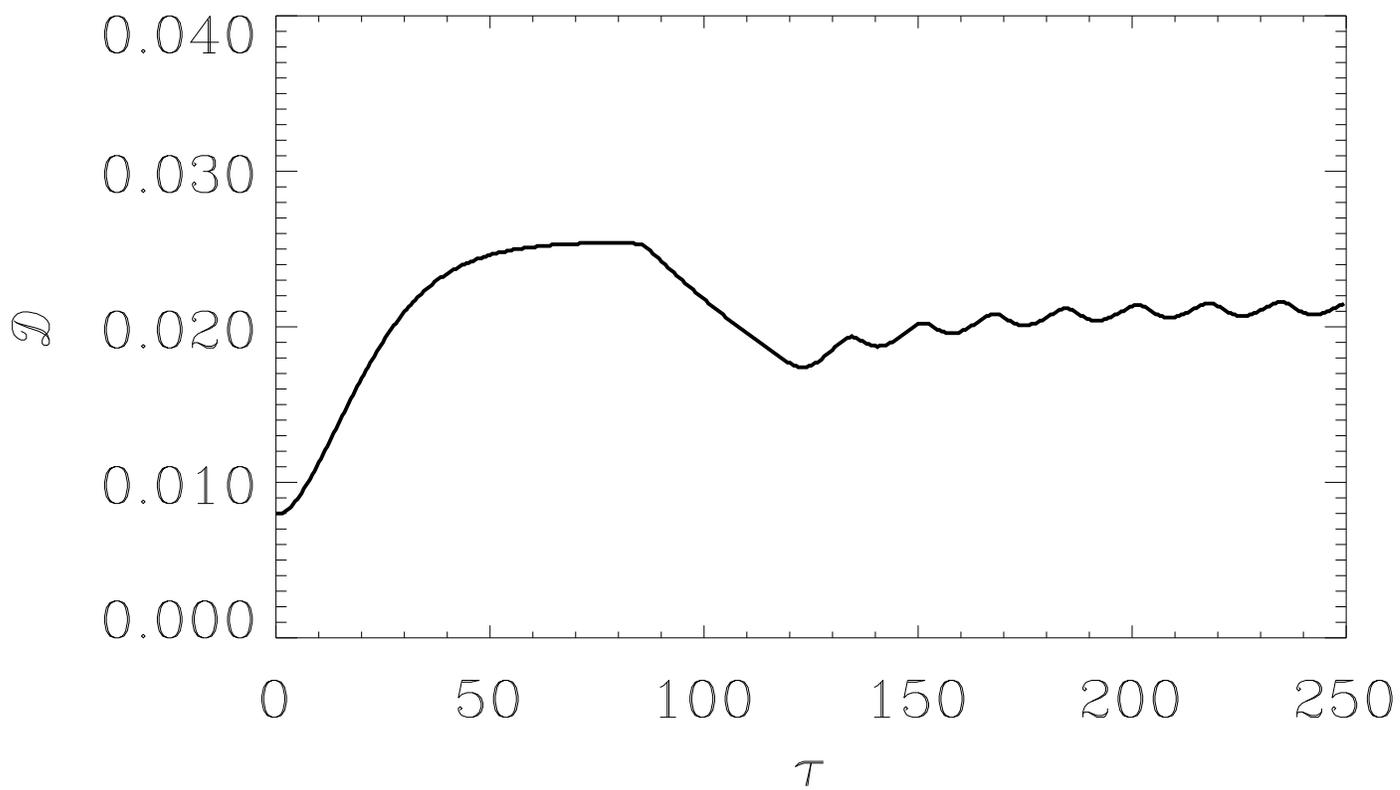
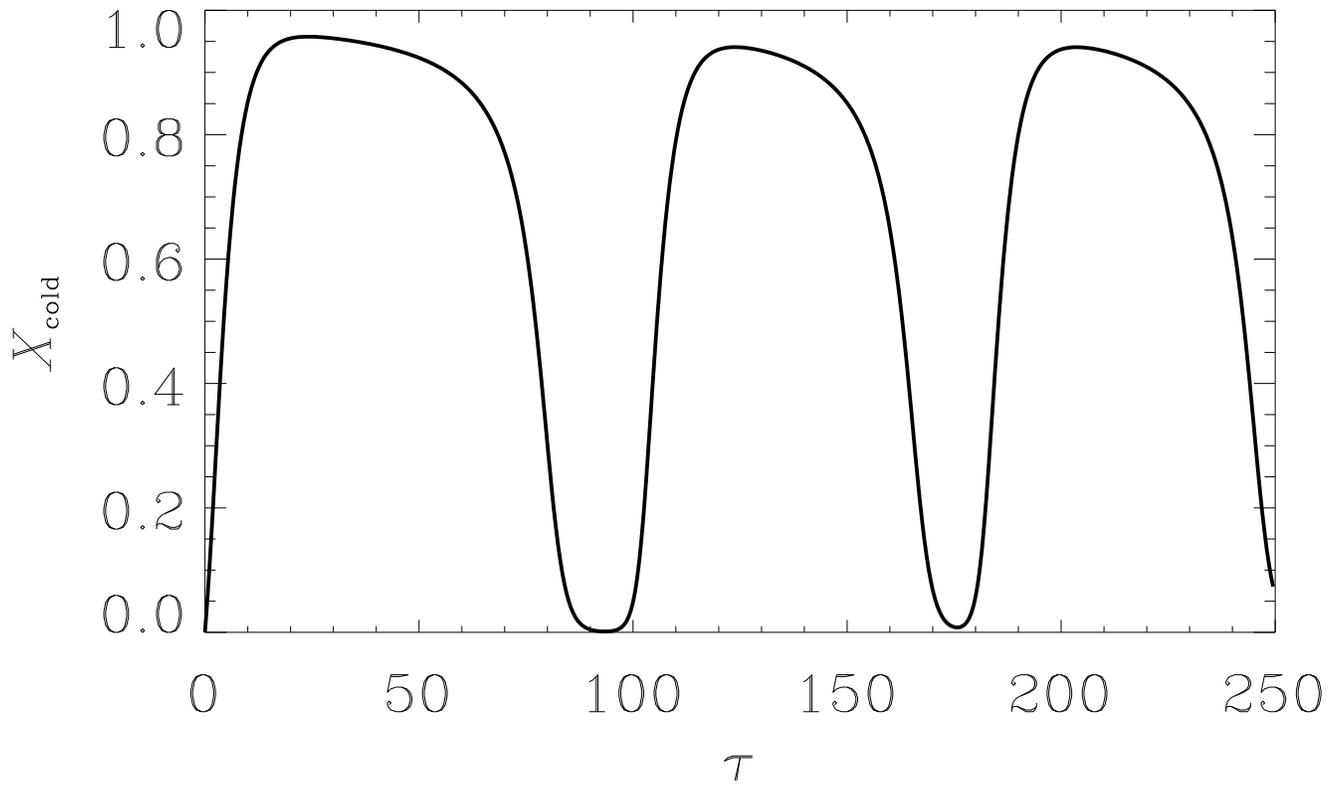
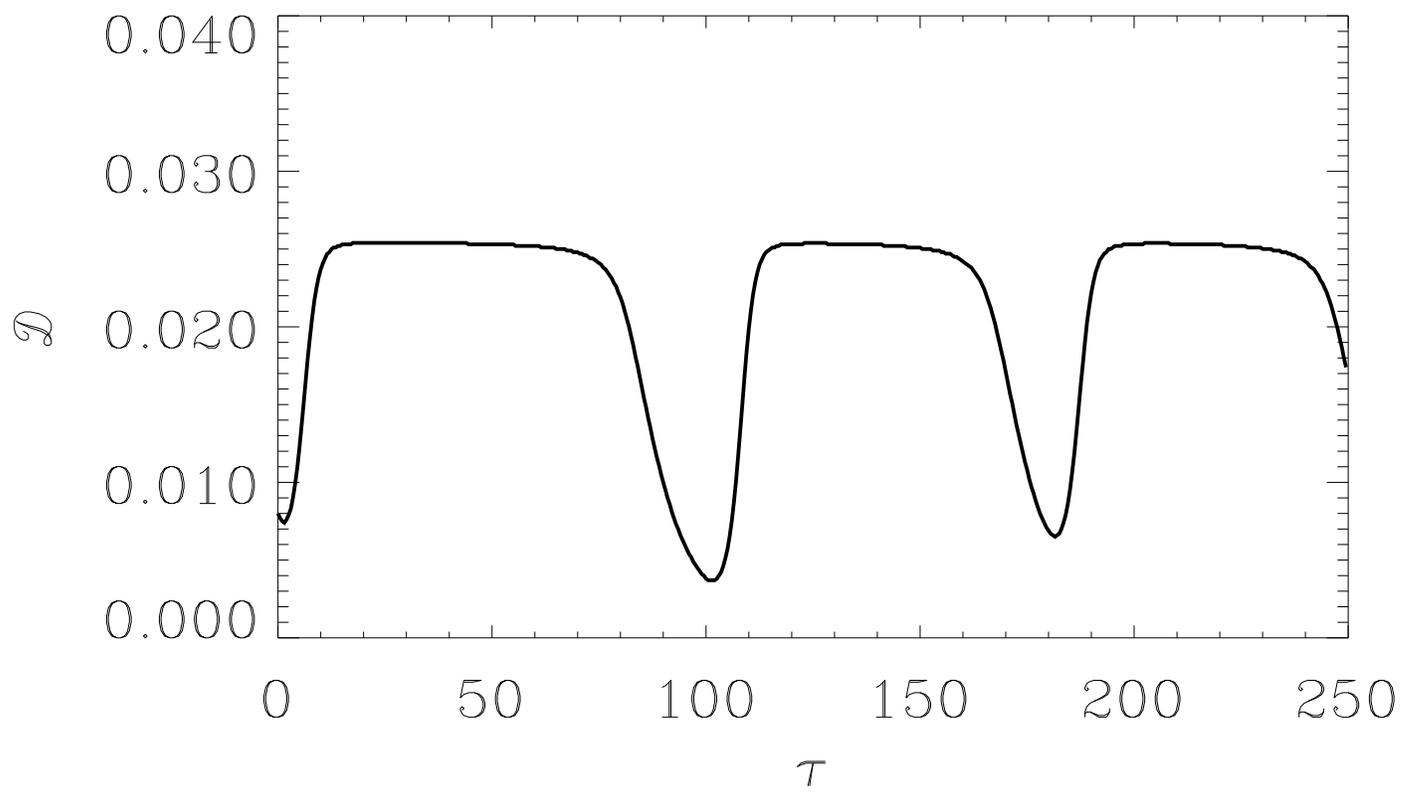
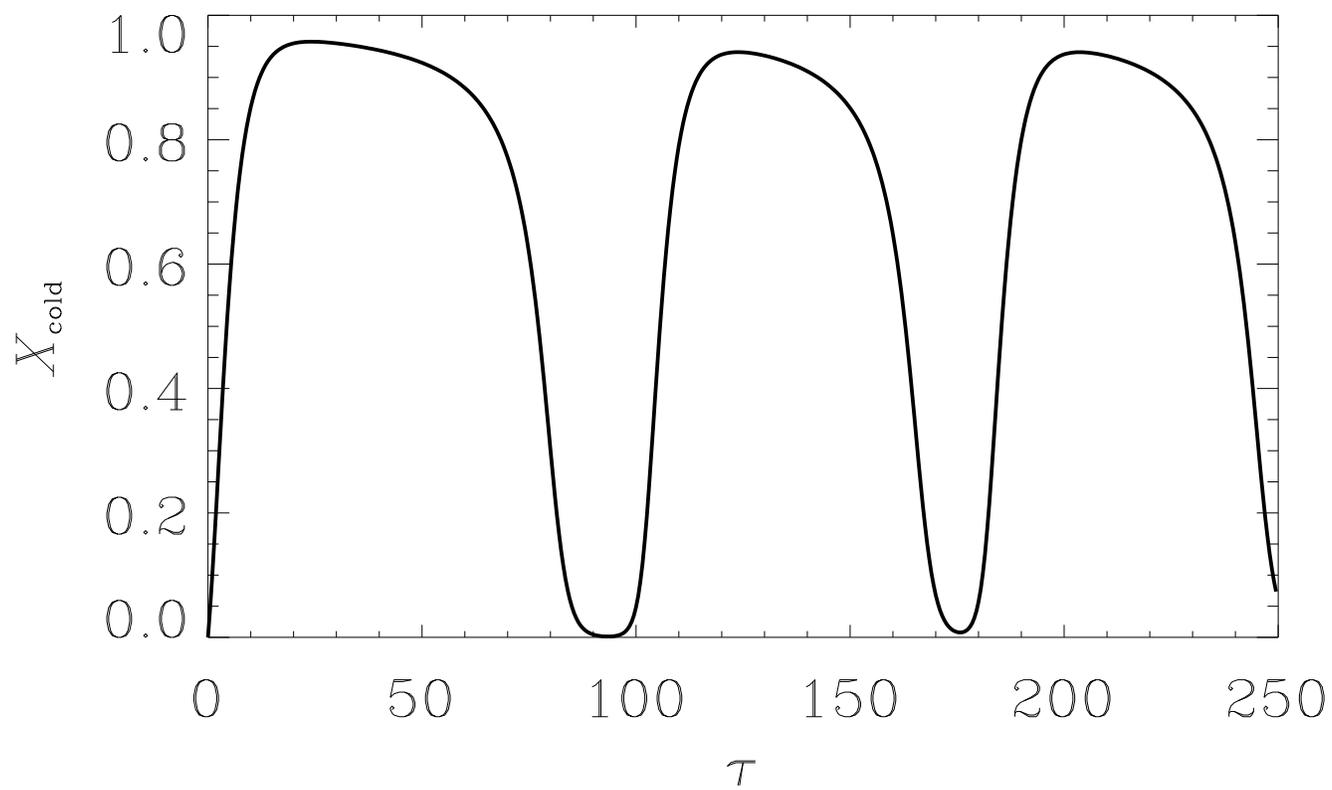
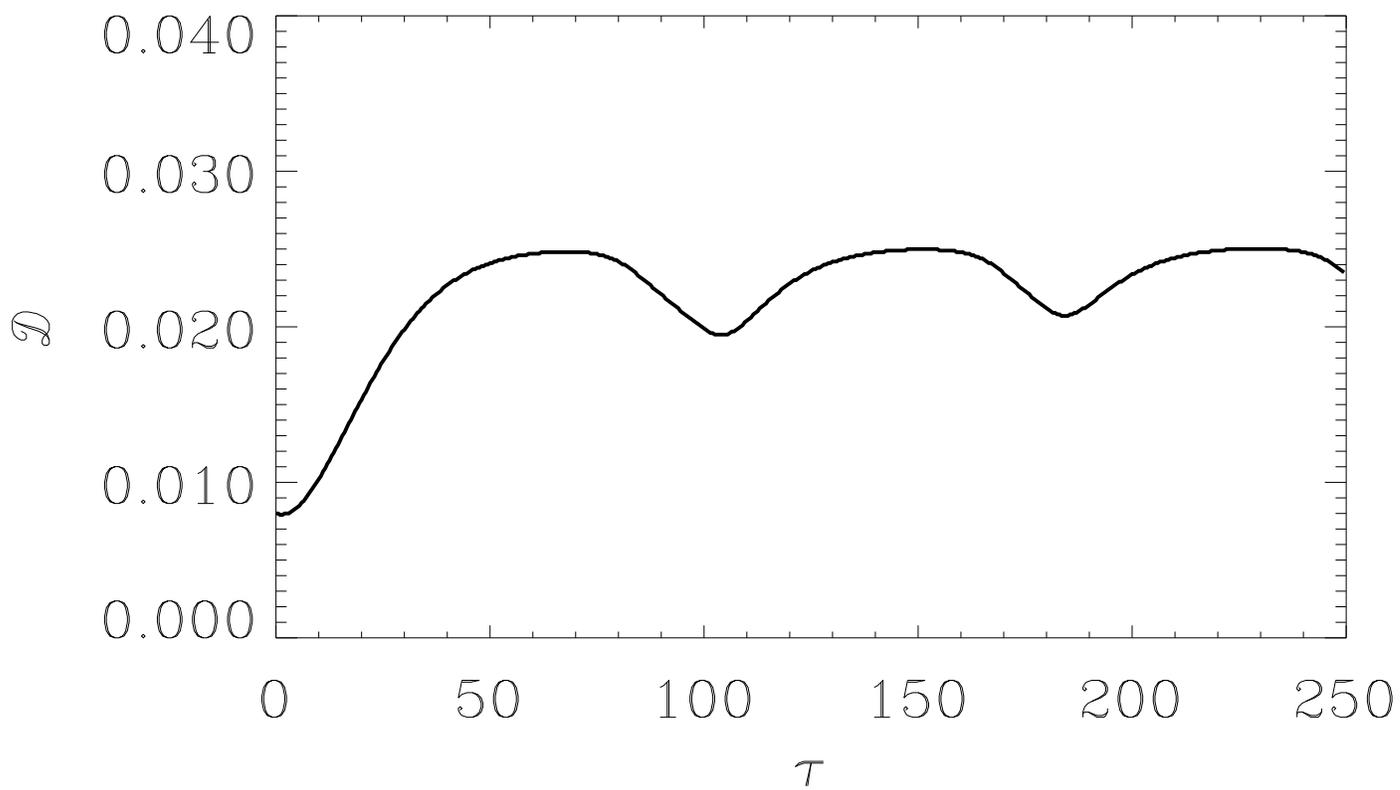
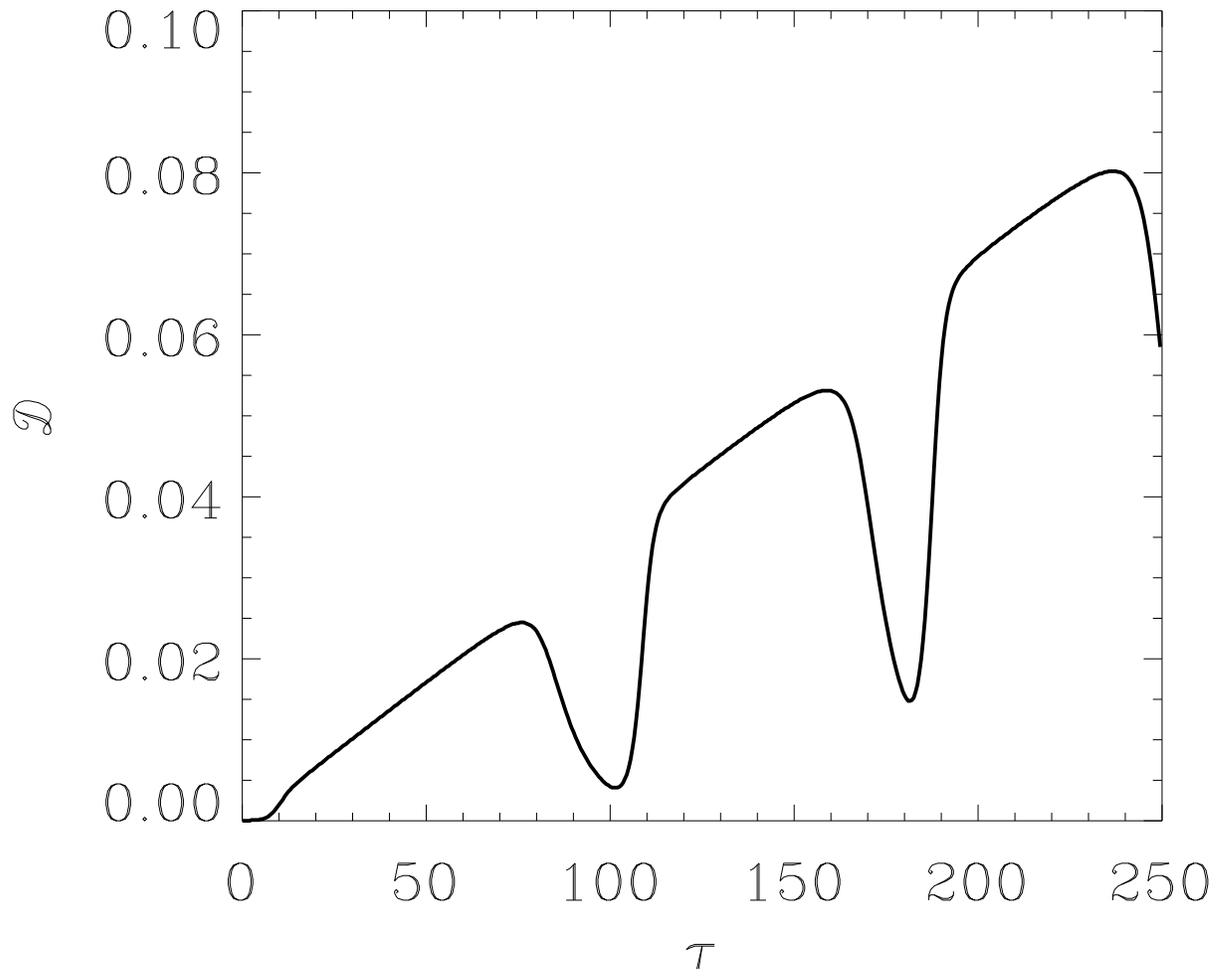
\end{document}